\documentclass[3p,times]{elsarticle}

\usepackage{ecrc}

\volume{00}

\firstpage{1}

\journalname{Nuclear Physics A}

\runauth{D. Stocco}

\jid{nupha}

\usepackage{amssymb}

\usepackage[figuresright]{rotating}

\newcommand{\pt}{\ensuremath{p_{\rm t}}}

\begin{document}

\begin{frontmatter}



\dochead{}

\title{Measurement of heavy-flavour decay muon production at forward rapidity in pp and Pb-Pb collisions at $\sqrt{s_{NN}} = 2.76$~TeV with the ALICE experiment}


\author{D. Stocco}
\ead{diego.stocco@subatech.in2p3.fr}
\author{for the ALICE Collaboration}

\address{SUBATECH (Ecole des Mines, CNRS-IN2P3, Universit\'e de Nantes), Nantes, France}

\begin{abstract}
The ALICE experiment measured the heavy-flavour production in the semi-muonic decay channel at forward rapidities ($2.5<y<4$) in pp and Pb--Pb collisions at $\sqrt{s_{{NN}}}~=~2.76$~TeV. We report on the first results on the \pt-differential cross-sections in pp collisions as well as on the nuclear modification factors as a function of the transverse momentum and centrality.
\end{abstract}


\end{frontmatter}


\section{Introduction}
The main goal of the ALICE experiment~\cite{Aamodt:2008zz} is the study of the properties of the state of strongly-interacting matter at very high energy density created in ultra-relativistic heavy-ion collisions at the LHC.
Heavy-flavour quarks (charm and beauty) have an important role in the investigation: being produced in the early stage of the collision, they are sensitive probes of the Quark-Gluon Plasma and allow us to study the parton-medium interaction.
The study of the spectra of heavy-flavoured hadrons provides information on the mechanisms of in-medium energy-loss and hadronization of heavy quarks.
The nuclear modification factor is a sensitive observable for this analysis.
It is defined as the ratio between the transverse momentum (\pt) spectra measured in ion--ion (A--A) collisions (corrected for the detector acceptance and response) and the \pt-differential cross-section measured in proton--proton (pp) collisions, rescaled by the nuclear overlap function estimated through the Glauber model~\cite{Miller:2007ri}:

\begin{equation}
R_{AA}(p_{\rm t}) = \frac{1}{\langle T_{AA} \rangle} \frac{\mathrm{d}N_{AA}/\mathrm{d}p_{\rm t}}{\mathrm{d}\sigma_{pp}/\mathrm{d}p_{\rm t}}
\end{equation}

The ALICE experiment has measured the nuclear modification factor of charmed mesons~\cite{Conesa:HP2012} and of heavy flavours in the semi-electronic~\cite{Kweon:HP2012} and semi-muonic decay channels.
The latter will be detailed in the following.

\section{Analysis}
The ALICE experiment is equipped with several detectors for tracking, particle identification, triggering and centrality estimation~\cite{Aamodt:2008zz}.
The most relevant detectors for the current analysis are the Silicon Pixel Detector (SPD) which covers the range $|\eta|<2$ and is used both for triggering and to measure the interaction vertex position; the VZERO, consisting of an array of scintillator hodoscopes covering the ranges $2.8<\eta<5.1$ and $-3.7<\eta<-1.7$, which is used for triggering and centrality estimation through a Glauber model fit of the signal amplitudes~\cite{Aamodt:2010cz}; and the Muon Spectrometer ($-4<\eta<-2.5$), consisting of a passive front absorber, five tracking stations (the central one placed inside a 3 T$\cdot$m integrated dipole magnetic field) and two trigger stations placed downstream of an iron filter, which is used to track and identify muons with momenta higher than 4 GeV/$c$.

The analysis was performed using data from pp and Pb--Pb collisions at $\sqrt{s_{NN}}$ = 2.76 TeV, collected in spring 2011 and fall 2010, respectively.
The data sample in Pb--Pb collisions consists of Minimum Bias (MB) events obtained requiring a signal in the SPD or in either of the two VZERO arrays in coincidence with the beam-beam counters.
In pp collisions, an additional muon trigger is used which requires, on top of the MB conditions, the detection of a muon with a transverse momentum above 0.5 GeV/$c$ by the muon trigger chambers.

Muons are identified by requiring that a track reconstructed in the tracking chambers matches a corresponding track segment in the trigger chambers.
This condition rejects most of the reconstructed hadrons, which are absorbed in the iron wall.
Geometrical cuts on the track pseudo-rapidity are then applied to remove the contamination of particles leaking into the spectrometer from outside the front absorber acceptance.
Moreover, the correlation between the track momentum and the distance of closest approach to the interaction point (DCA\footnote{The DCA is defined as the distance between the interaction point and the extrapolation of the track to the plane orthogonal to the beam-line and containing the interaction point itself.})  is used to remove fake tracks and tracks from beam-gas interactions.
The main background contribution after selection cuts consists of muons from the decay in flight of light hadrons produced in the collision.
In the pp analysis, such contribution was estimated through Monte Carlo simulations, using the Phojet and Pythia event generators as input, and then subtracted from the measured inclusive spectrum (see~\cite{Abelev:2012pi} and references therein for details).
This approach, however, could not be used in the Pb--Pb analysis, due to the presence of unknown nuclear effects, in particular medium-induced parton energy loss at forward rapidity.
Hence, a data driven method was developed, based on the pion and kaon distributions measured in pp and Pb--Pb collisions in the ALICE central barrel.
Such distributions are extrapolated to forward rapidities and used to generate the corresponding decay muons through simulations of the decay kinematics and the front absorber (see~\cite{Abelev:2012qh} for further details).
The background contribution decreases with increasing transverse momentum: the systematic uncertainty on its subtraction can therefore be limited by restricting to $\pt>2$ (4)~GeV/$c$ in pp (Pb--Pb) collisions.

\begin{figure}[!ht]
 \centering
 \includegraphics[width=0.45\textwidth]{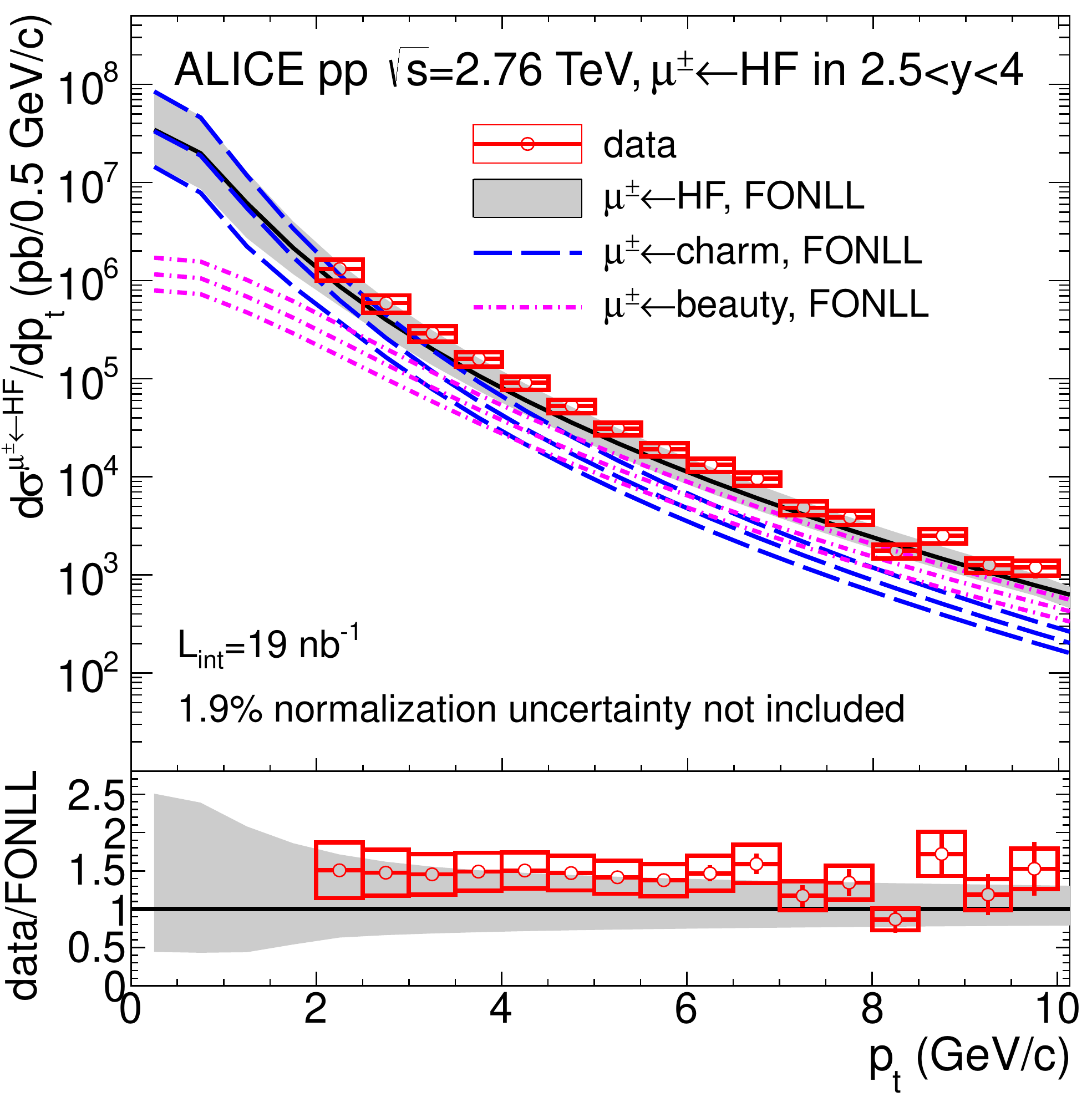}
 \caption{Transverse momentum differential cross-section of muons from heavy-flavour decay in pp collisions at $\sqrt{s}$ = 2.76~TeV (red open circles). The vertical bars (boxes) are the statistical (systematic) uncertainties. The data points are compared to the sum (grey band) of the contribution of muons from charm (blue dashes) and beauty (magenta dash-dotted) decays, estimated with FONLL calculations~\cite{Cacciari:1998it,Cacciari:2012ny}. The ratio between data and FONLL is shown in the bottom panel.}
 \label{fig:ppRef}
\end{figure}

The \pt-differential cross-section of muons from heavy-flavour decays, measured in pp collisions at $\sqrt{s}$ = 2.76~TeV with an integrated luminosity of $\mathcal{L}_{int} = 19$~nb$^{-1}$, is shown in Figure~\ref{fig:ppRef}.
The vertical bars are the statistical uncertainties.
The boxes are the uncorrelated systematic uncertainties, accounting for detection efficiencies, alignment and background subtraction.
The results are compared to Fixed-Order Next-to-Leading Log (FONLL) calculations~\cite{Cacciari:1998it,Cacciari:2012ny}, which show a good agreement with data within errors.

\begin{figure}[!ht]
 \centering
\includegraphics[width=0.75\textwidth,trim=0 0 30 0,clip]{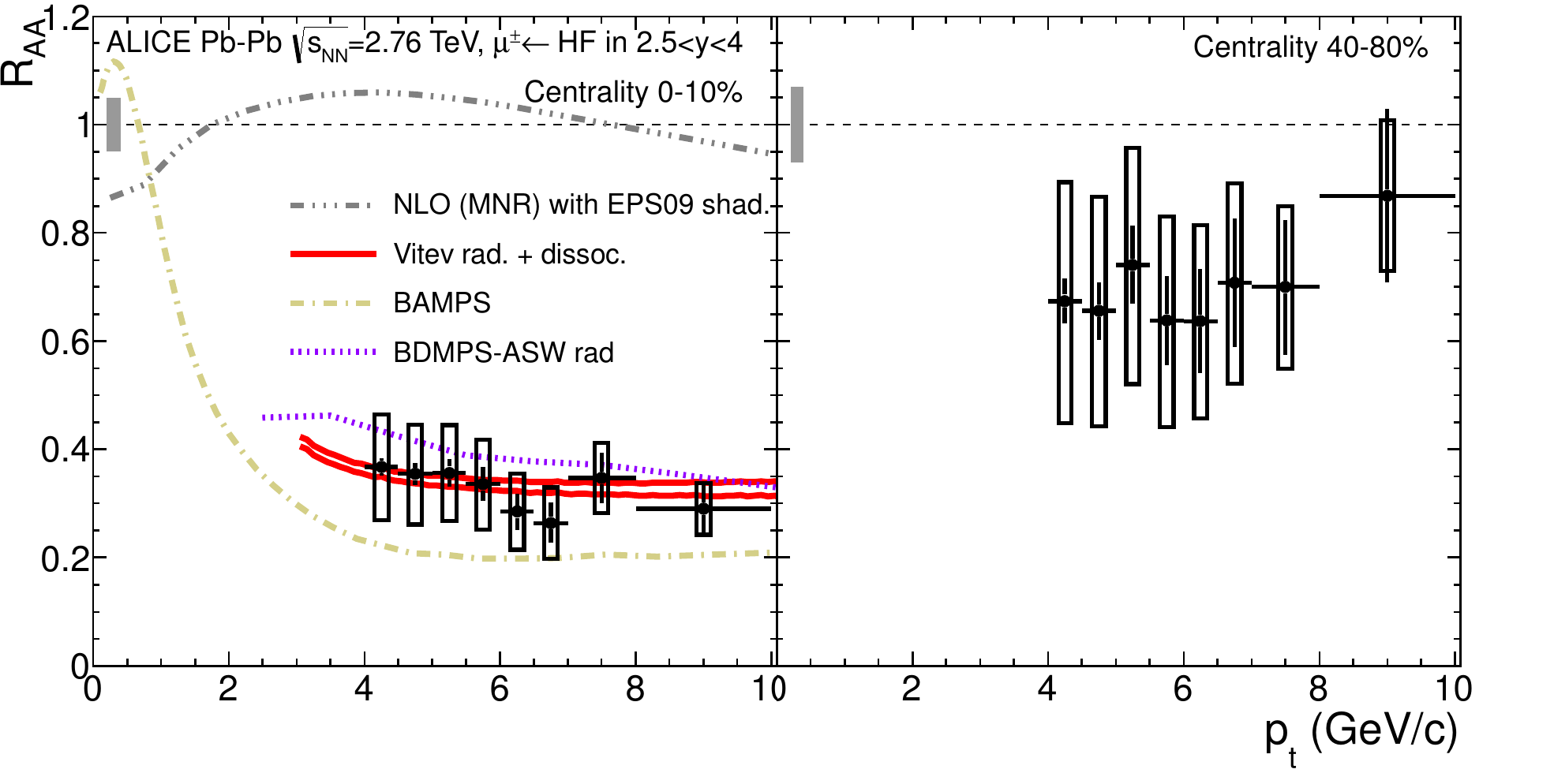}
 \caption{Nuclear modification factor as a function of transverse momentum for muons from heavy-flavour decay (red circles) in the 0-10\% central (left) and 40-80\% peripheral (right) collisions. The vertical bars (boxes) are the statistical (uncorrelated systematic) uncertainties. The grey box at 1 is the correlated error on the centrality estimation and the pp cross-section normalization. The expected contribution of shadowing, estimated using perturbative calculations by Mangano, Nason and Ridolfi~\cite{Mangano:1991jk} and the EPS09NLO~\cite{Eskola:2009uj} parameterization, is also shown (dot-dot-dot-dashed curve). The $R_{AA} (\pt)$ in the most central collisions is compared with models implementing collisional (BAMPS)~\cite{Uphoff:2012gb} radiative (BDMPS-ASW)~\cite{Armesto:2005iq} and radiative with in-medium hadronization (Vitev \textit{et al.})~\cite{Sharma:2009hn} energy-loss.}
 \label{fig:raaVsPt}
\end{figure}

Figure~\ref{fig:raaVsPt} shows the nuclear modification factor as a function of the transverse momentum for muons from heavy-flavour decay in the 0--10\% (left panel) and 40--80\% (right panel) most central collisions.
The vertical bars (boxes) are the statistical (uncorrelated systematic) uncertainties.
The correlated uncertainties on $\langle T_{AA} \rangle$ and on the cross-section normalization of the pp reference are shown as a filled box at $R_{AA} = 1$.
The nuclear modification factor is independent of $\pt$ within uncertainties, and exhibits a reduction of a factor 3--4 in the most central collisions.
It is worth noting that the in-medium energy loss is not the only mechanism that could lead to a reduction of the $R_{AA}$.
In particular, the nuclear modification of the parton distributions in nuclei could lead to a variation of the initial hard-scattering probability, and a consequent variation of the heavy-flavour yield.
In the kinematic range relevant for heavy-flavour production the main effect is the nuclear shadowing, which reduces the parton distribution functions for partons carrying a fraction of the nucleon momentum smaller than $10^{-2}$.
This effect was estimated by using perturbative calculations by Mangano, Nason and Ridolfi~\cite{Mangano:1991jk} and the EPS09NLO~\cite{Eskola:2009uj} parameterization of the shadowing.
The result is shown in Figure~\ref{fig:raaVsPt} (grey dot-dot-dot-dashed curve): the initial state effect is expected to be small in the {\pt} region studied, thus suggesting that the strong reduction is a final-state effect.
The $R_{AA}$ in the most central collisions (left panel) is compared to models implementing collisional (BAMPS)~\cite{Uphoff:2012gb}, radiative (BDMPS-APW)~\cite{Armesto:2005iq} and radiative with in-medium hadronization (Vitev \textit{et al.)}~\cite{Sharma:2009hn} in-medium energy loss: a good agreement with data is found for the last two models, while the BAMPS tends to underestimate the heavy-flavour muons $R_{AA}$.
It is worth noting that the comparison of these models with the $R_{AA}$ of D mesons measured at mid-rapidity with ALICE~\cite{ALICE:2012ab} leads to similar observations.

\begin{figure}[!ht]
 \centering
 \includegraphics[width=0.45\textwidth]{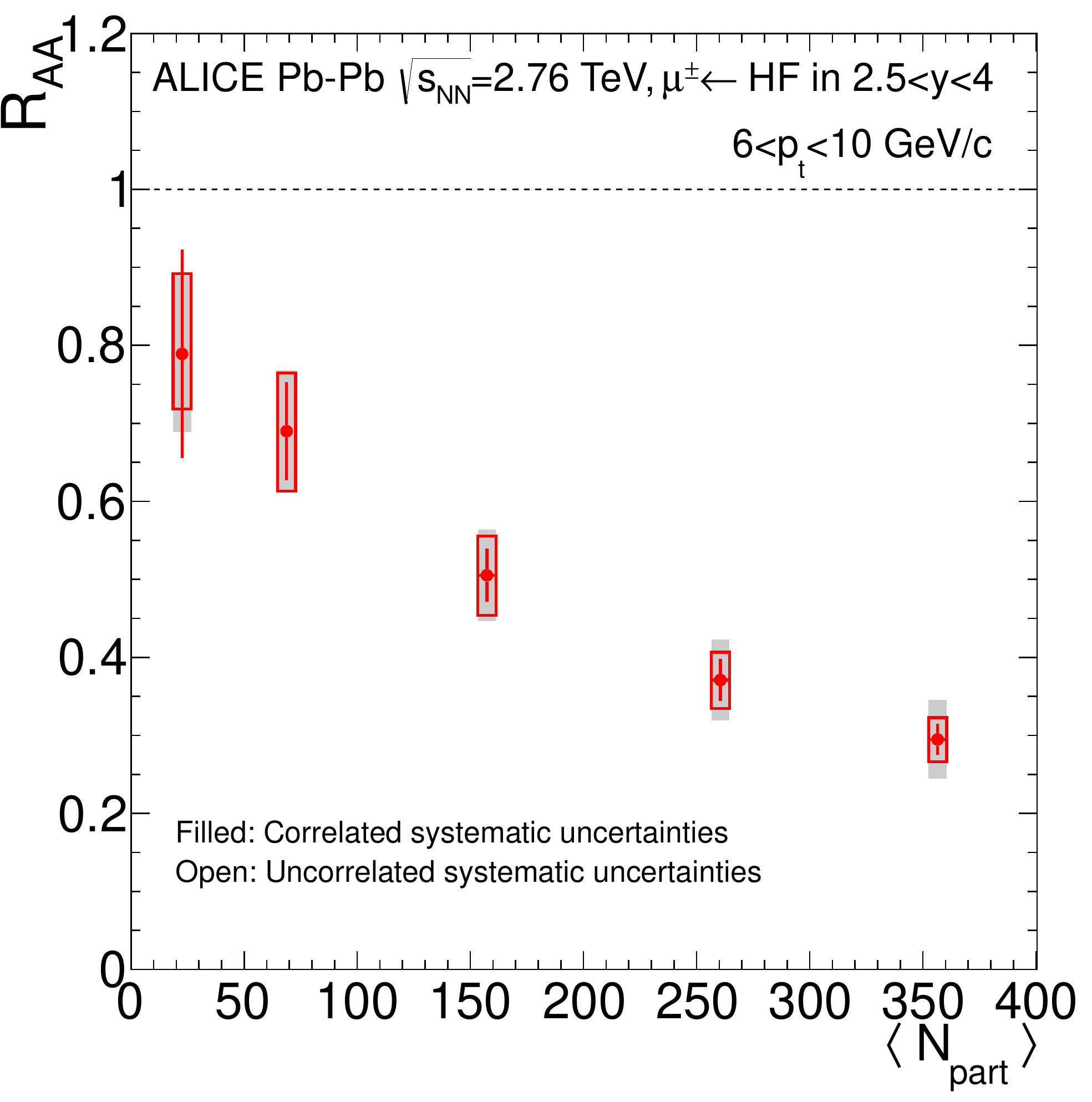}
 \caption{Nuclear modification factor as a function of centrality for muons from heavy-flavour decay with $\pt > 6$~GeV/$c$. The vertical bars are the statistical uncertainties. The empty (filled) boxes are the uncorrelated (correlated) systematic uncertainties.}
  \label{fig:raaVsCentrality}
\end{figure}

Finally, the centrality dependence of the nuclear modification factor is shown in Figure~\ref{fig:raaVsCentrality}.
The vertical bars (boxes) are the statistical (uncorrelated systematic) uncertainties, while the grey filled boxes are the systematic uncertainties on the nuclear overlap function and on the normalization of the pp reference.
The result refers to muons from heavy-flavour decay with a transverse momentum higher than 6 GeV/$c$.
The high $\pt$ cut allows the selection of a region dominated by beauty decay according to the FONLL predictions (see also Figure~\ref{fig:ppRef}).
This nuclear modification factor is similar to the one of D mesons measured at mid-rapidity in $6<\pt<12$~GeV/$c$ with ALICE~\cite{ALICE:2012ab} and of non-prompt J/$\psi$ measured in $|y|<2.4$ and $6.5<\pt<30$~GeV/$c$ by the CMS Collaboration~\cite{Chatrchyan:2012np}.

\section{Conclusion}
The ALICE experiment has measured the heavy-flavour production in the semi-muonic decay channels at forward rapidities ($2.5<y<4$) in pp and Pb--Pb collisions at 2.76~TeV center of mass energy.
The measured \pt-differential cross section is well described by the FONLL perturbative QCD calculations.

The resulting nuclear modification factor as a function of $\pt$ (for $\pt>4$~GeV/$c$) was also shown, for the 0-10\% most central and the 40-80\% most peripheral collisions.
A reduction of a factor of 3--4 is observed in the most central collisions, independent of $\pt$.
It is worth noting that the initial state effects are expected to be small in this transverse momentum region.
A similar reduction is observed when measuring the nuclear modification factor as a function of centrality for muons from heavy-flavour decay with $\pt>6$~GeV/$c$, a region where the beauty decay contribution is expected to be dominant according to FONLL calculations.





\bibliographystyle{elsarticle-num}
\bibliography{biblio}

\begin{thebibliography}{10}
\expandafter\ifx\csname url\endcsname\relax
  \def\url#1{\texttt{#1}}\fi
\expandafter\ifx\csname urlprefix\endcsname\relax\def\urlprefix{URL }\fi
\expandafter\ifx\csname href\endcsname\relax
  \def\href#1#2{#2} \def\path#1{#1}\fi

\bibitem{Aamodt:2008zz}
K.~Aamodt, et~al., JINST 3 (2008) S08002.

\bibitem{Miller:2007ri}
M.~L. Miller, K.~Reygers, S.~J. Sanders, P.~Steinberg, Ann.Rev.Nucl.Part.Sci.
  57 (2007) 205--243.
\newblock \href {http://arxiv.org/abs/nucl-ex/0701025}
  {\path{arXiv:nucl-ex/0701025}}.

\bibitem{Conesa:HP2012}
Z.~Conesa~del Valle, These proceedings.

\bibitem{Kweon:HP2012}
M.-J. Kweon, These proceedings.

\bibitem{Aamodt:2010cz}
K.~Aamodt, et~al., Phys.Rev.Lett. 106 (2011) 032301.
\newblock \href {http://arxiv.org/abs/1012.1657} {\path{arXiv:1012.1657}}.

\bibitem{Abelev:2012pi}
B.~Abelev, et~al., Phys.Lett. B708 (2012) 265--275.
\newblock \href {http://arxiv.org/abs/1201.3791} {\path{arXiv:1201.3791}}.

\bibitem{Abelev:2012qh}
B.~Abelev, et~al.{Submitted to PRL}.
\newblock \href {http://arxiv.org/abs/1205.6443} {\path{arXiv:1205.6443}}.

\bibitem{Cacciari:1998it}
M.~Cacciari, M.~Greco, P.~Nason, JHEP 9805 (1998) 007.
\newblock \href {http://arxiv.org/abs/hep-ph/9803400}
  {\path{arXiv:hep-ph/9803400}}.

\bibitem{Cacciari:2012ny}
M.~Cacciari, S.~Frixione, N.~Houdeau, M.~L. Mangano, P.~Nason, G.~Ridolfi,
\newblock \href{http://arxiv.org/abs/1205.6344}
  {\path{arXiv:1205.6344}}.

\bibitem{Mangano:1991jk}
M.~L. Mangano, P.~Nason, G.~Ridolfi, Nucl.Phys. B373 (1992) 295--345.

\bibitem{Eskola:2009uj}
K.~Eskola, H.~Paukkunen, C.~Salgado, JHEP 0904 (2009) 065.
\newblock \href {http://arxiv.org/abs/0902.4154} {\path{arXiv:0902.4154}}.

\bibitem{Uphoff:2012gb}
J.~Uphoff, O.~Fochler, Z.~Xu, C.~Greiner,
\newblock \href {http://arxiv.org/abs/1205.4945}
  {\path{arXiv:1205.4945}}.

\bibitem{Armesto:2005iq}
N.~Armesto, A.~Dainese, C.~A. Salgado, U.~A. Wiedemann, Phys.Rev. D71 (2005)
  054027.
\newblock \href {http://arxiv.org/abs/hep-ph/0501225}
  {\path{arXiv:hep-ph/0501225}}.

\bibitem{Sharma:2009hn}
R.~Sharma, I.~Vitev, B.-W. Zhang, Phys.Rev. C80 (2009) 054902.
\newblock \href {http://arxiv.org/abs/0904.0032} {\path{arXiv:0904.0032}}.

\bibitem{ALICE:2012ab}
B.~Abelev, et~al.{Submitted to JHEP}.
\newblock \href {http://arxiv.org/abs/1203.2160} {\path{arXiv:1203.2160}}.

\bibitem{Chatrchyan:2012np}
S.~Chatrchyan, et~al., JHEP 1205 (2012) 063.
\newblock \href {http://arxiv.org/abs/1201.5069} {\path{arXiv:1201.5069}}.

\end{thebibliography}







\end{document}